\documentclass[12pt,a4paper,final]{iopart}

\usepackage{iopams}  
\usepackage[breaklinks=true,colorlinks=true,linkcolor=blue,urlcolor=blue,citecolor=blue]{hyperref}
\usepackage[pdftex]{graphicx}
 \usepackage{tikz}
\usetikzlibrary{decorations.markings}
\usepackage{scrextend}
\def\beq{\begin{equation}}
\def\eeq{\end{equation}}
\def\bsp{\begin{split}}
\def\esp{\end{split}}
\def\bea{\begin{eqnarray}}
\def\eea{\end{eqnarray}}
\def\ba{\begin{array}}
\def\ea{\end{array}}

\def\dg{\dagger}

\def\lb{\left(}
\def\rb{\right)}

\def\l.{\left.}
\def\r.{\right.}

\def\ra{\rangle}
\def\la{\langle}

\def\bo{\bold{k}}

\begin{document}

\title[Quantum Kagom\'e ice hardcore bosons]{Ground state properties of  quantum Kagom\'e ice hardcore bosons}

\author{S. A. Owerre $^{1,2}$}
\address{$^1$ African Institute for Mathematical Sciences, 6 Melrose Road, Muizenberg, Cape Town 7945, South Africa.}
\address{$^2$Perimeter Institute for Theoretical Physics, 31 Caroline St. N., Waterloo, Ontario N2L 2Y5, Canada.}
\ead{sowerre@perimeterinstitute.ca}

%

\begin{abstract}
We study the quantum Kagom\'e ice hardcore bosons, which corresponds to the XY limit of quantum spin ice Hamiltonian.   We  estimate the values of the zero-temperature thermodynamic quantities using the large-$S$ expansion. We show that our semi-classical analysis is consistent   with finite temperature  quantum Monte Carlo estimates.   
\end{abstract}

\pacs{75.10.Jm, 05.30.Jp}
\vspace{2pc}
\noindent{\textit Keywords}: hardcore bosons, condensate fraction, dynamical structure factors.

\submitto{\  J. Phys. A: Math. Theor.}

\section{Introduction}
The  recent search for a putative  quantum spin liquid (QSL) state in three-dimensional (3D)  pyrochlore lattice  quantum spin ice (QSI) has led to a new proposal of quantum spin Hamiltonian  devoid of the debilitating quantum Monte Carlo (QMC) sign problem in a wide range of the parameter regime \cite{ Huang, gin,juan}. Recently,   Carrasquilla {\it et~al.} \cite{juan},  have studied the projected   3D QSI Hamiltonian of Huang, Chen, and Hermele \cite{Huang} onto a 2D frustrated  Kagom\'e lattice with a [111] crystallographic field. They have explored this model by an explicit finite temperature QMC simulations and identified the interactions that promote a QSL state. In a recent study, we have complemented the QMC analysis using linear spin wave theory \cite{sow1}. In that study, we found that quantum fluctuations are suppressed in the quantum Kagom\'e ice model, hence the semiclassical analysis captures a broad trend of the quantum phase diagram and ground state properties uncovered by QMC \cite{juan}.

However, there remains the possibility of the XY limit of the quantum Kagom\'e ice (QKI) model.   The XY model essentially controls the dynamics of the fully frustrated system, \cite{ Huang,juan} and it forms the basis of ultra-cold atoms trapped in quantum optical lattices \cite{mar,duan,st}.  The quantum  XY model is also  known to describe quantum  magnetic insulators and quantum Hall bilayer systems \cite{kry}. Thus, it is crucial to understand the nature of the XY limit of this model.   We consider the hardcore boson model
\begin{eqnarray}
\mathcal{H}=-t\sum_{\la lm\ra}\lb b_{l}^\dagger b_{m} + b_{m}^\dagger b_{l}\rb-t^\prime\sum_{\la lm\ra}\lb b_{l}^\dagger b_{m}^\dagger + b_{m} b_{l}\rb -\mu\sum_l n_l,
\end{eqnarray}
where $t$ and $t^\prime$ are the hopping amplitudes between neighbouring sites, $\mu$ is a chemical potential, $b_{l}^\dagger( b_{l})$  are the bosonic creation (annihilation) operators at site $l$, and $n_l= b_{l}^\dagger b_{l}$ is the occupation number. This Hamiltonian  maps to XY model via the Matsubara-Matsuda transformation \cite{mat}, $S_l^+ \to b^\dagger_l,~S_l^-\to b_l,~S_l^z\to n_l-1/2$,
 \begin{eqnarray}
&\mathcal{H}=-\sum_{\la lm\ra}{\mathcal{J}_{\pm\pm}}\lb S_{l}^+S_{m}^+ + S_{l}^-S_{m}^-\rb+{\mathcal{J}_\pm}\lb S_{l}^+S_{m}^- +S_{l}^-S_{m}^+\rb\label{k2} -H_z\sum_{l}  S_l^z,
\end{eqnarray}
where $S^{\pm}=S^x\pm iS^y$ are operators that flip the spins at site $l$,  $t\to\mathcal{J}_{\pm},~ t^\prime \to \mathcal{J}_{\pm\pm}$, and $\mu\to H_z$. 
The  Hamiltonian \ref{k2} corresponds to the XY limit of  QSI model  \cite{Huang, juan}.  A crucial observation in this model is that the sign of $\mathcal{J}_{\pm\pm}$ is irrelevant. It can be changed by a $\pi/2$-rotation about the $z$-axis: $S_{l,m}^{\pm}\to\pm iS_{l,m}^{\pm}$, leaving  the other terms invariant. Hence, the ground state of Eq.~\ref{k2} is independent of the   sign of $\mathcal{J}_{\pm\pm}$ but depends on the sign of $\mathcal{J}_{\pm}$ for non-bipartite lattices.  We consider  $\mathcal{J}_{\pm}>0$.   The usual  hard-core bosons  is recovered when  $\mathcal{J}_{\pm\pm}= 0$ \cite{ber, tom, B,san, fu,fu1, gom, zhe,zhe1}. The effects of dominant $\mathcal{J}_{\pm\pm}$ is well-pronounced only on non-bipartite lattices, and it has not been reported in the existing literatures on the Kagom\'e lattice. 
 
The goal of this paper is to provide the estimated values of the thermodynamic  quantities for the quantum Kagom\'e ice hardcore bosons in the dominant $t^\prime$ limit. For this purpose  we study the nature of Eq.~\ref{k2}, focusing mainly on large  $\mathcal{J}_{\pm\pm}$ limit. Henceforth, we take $\mathcal{J}_{\pm\pm}$ as the energy unit and consider the Hamiltonian
\begin{eqnarray}
\mathcal{H}&=-\frac{1}{2}\sum_{\la lm\ra}\bigg[\lb S_{l}^+S_{m}^+ + S_{l}^-S_{m}^-\rb\label{k1}+{\Delta}\lb S_{l}^+S_{m}^- + S_{l}^-S_{m}^+\rb\bigg]\nonumber\\&
-H_x\sum_l S_l^x-H_z\sum_l S_l^z,
\end{eqnarray}
 where $0\leq \Delta= \mathcal{J}_{\pm}/\mathcal{J}_{\pm\pm}\leq 1$.  The external magnetic fields are introduced to enable the calculation of magnetizations and susceptibilities.  The Hamiltonian \ref{k1}, retains  $Z_2$-invariance in the $x$-$y$ plane when $H_x=0$, that is a $\pi$-rotation about the $z$-axis in spin space: $S_{l,m}^z\to S_{l,m}^z$,  $S_{l,m}^\pm\to -S_{l,m}^\pm$.   At $H_{x,z}=0$, the limits $\Delta=0$ and $\Delta=1$ correspond to the isotropic $Z_2$-invariant XY model and the fully polarized $S_x$ (Ising) ferromagnet respectively. We present an analysis based on the semiclassical large-$S$ expansion and we show that the estimated values of our analysis are in good agreement with finite temperature QMC simulation by Carrasquilla \footnote{\label{juan1} The QMC results were provided by J.  Carrasquilla from his analysis of the fully frustrated system in Ref.~\cite{juan}.} on the Kagom\'e lattice.  

\begin{figure}[!]
\centering
\includegraphics[width=5in]{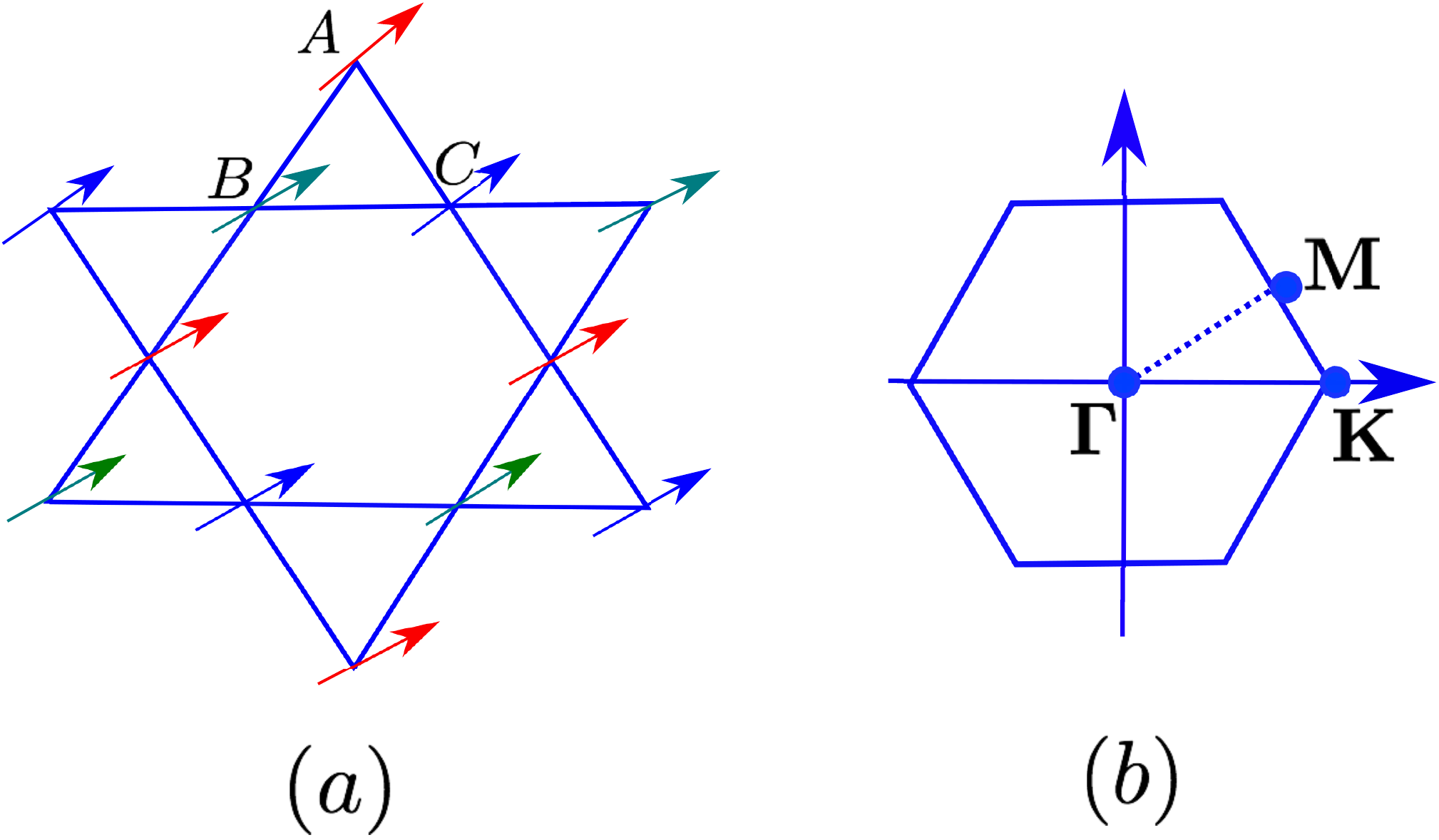}
\caption{Color online. (a) The three-sublattice ($A,~B~,C$) Kagom\'e lattice with canted ferromagnet.  (b). Brillouin zone of the Kagom\'e lattice indicating the high symmetry paths  ${\bf \Gamma}= (0,0)$; $\bold K=(2\pi/3, 0)$; $\bold M=(\pi/2, \pi/2\sqrt{3})$.}
\label{lattice}
\end{figure} 
\section{  Linear spin wave theory}
\subsection{  Mean-field theory} In the  mean-field analysis or large-$S$ limit, the spin operators in Eq.~\ref{k1} are replaced with classical vectors: $\bold{S}_l= S\bold{n}_l$, where $\bold{n}_l=\lb\sin\theta_l\cos\phi_l, \sin\theta_l\sin\phi_l,\cos\theta_l \rb$ is a unit vector. For this model, there is only one  possible phase at zero magnetic fields --- an  easy-axis ferromagnet (FM) resulting from the spontaneously broken $Z_2$ symmetry along the $x$-direction. This phase becomes a canted ferromagnet (CFM) for small $H_z$. For large $H_z$, there is a fully polarized (FP) ferromagnet, with the spins aligned along the $z$-axis. Both ferromagnets are described with $\theta_l=\theta$ and  $\phi_l=0$, hence the classical energy is given by

\begin{eqnarray}
e_c&=  -\zeta\lb 1+ \Delta \rb\sin^2\theta-h_{x}\sin\theta - h_{z}\cos\theta,
\label{mf1} 
\end{eqnarray}
where $e_c=\mathcal{E}_c/\mathcal{N}S^2$, $\mathcal{N}$ is the total number of sites, and  $\zeta=2,3$ on the Kagom\'e and the triangular lattices respectively, and $h_{x,z}=H_{x,z}/S$. 
 Minimizing the classical energy with respect to $\theta$ yields
\begin{eqnarray}
h_x= \tan\theta_c[h_z-2\zeta\lb 1+ \Delta \rb \cos\theta_c],
\label{hh}
\end{eqnarray}
hence
\begin{eqnarray}
e_c(\theta_c)&=  \zeta\lb 1+ \Delta \rb\sin^2\theta_c-h_z\sec\theta_c,
\label{mf2} 
\end{eqnarray}
where $\theta_c$ is a function of $h_x$. At $h_x= 0$, we get $h_z=h_F^c\cos\vartheta_c$ and the corresponding energy is $e_c(\theta_c=\vartheta_c)=-\zeta\lb 1+ \Delta \rb(1+\cos^2\vartheta_c)$, where $h_F^c=2\zeta(1+\Delta)$ is the critical field between the CFM and the FP. For $h_x\ll 1$, the energy is obtained perturbatively in $h_x$:

\begin{eqnarray}
e_c=e_c(h_x=0)+ h_x\left.\frac{\partial e_c(\theta_c)}{\partial h_x}\right\vert_{h_x=0} +\cdots,
\end{eqnarray}
where $e_c(h_x=0)=e_c(\theta_c=\vartheta_c)$.
\subsection{Holstein-Primakoff transformation}
We perform  linear spin wave theory (LSWT)  by rotating the coordinate about the $y$-axis so that the $z$-axis coincides with the local direction of the classical polarization. The corresponding rotation matrix is 
\begin{eqnarray}
 \mathcal{R}_y(\theta)= \left(
\begin{array}{ccc}
  \cos\theta& 0& \sin\theta \\
0& 1&0 \\
-\sin\theta & 0& \cos\theta \\  
 \end{array}
\right).
 \label{rot}
\end{eqnarray}
Next, we employ a three-sublattice Holstein-Primakoff transformation \cite{HP} with the bosonic creation and annihilation operators, $a^\dg_{l\alpha}$ and $ a_{l\alpha}$ \cite{jon} respectively, where $\alpha =A,~B,~C$ labels the three sublattices on the Kagom\'e lattice as depicted in Fig.~\ref{lattice}(a). After Fourier transform, 
the linearized momentum space Hamiltonian is given by\begin{eqnarray}
&\mathcal H= S\sum_{\bo,\alpha,\beta}\lb \mathcal{M}_{\alpha\beta}^0\delta_{\alpha\beta} +\mathcal{M}_{\alpha\beta}^-\rb \lb a_{\bo \alpha}^\dagger a_{\bo \beta}+a_{-\bo \alpha}^\dagger a_{-\bo \beta}\rb\label{main}\\&\nonumber +\mathcal{M}_{\alpha\beta}^+ \lb a_{\bo \alpha}^\dagger a_{-\bo \beta}^\dagger +a_{-\bo \alpha} a_{\bo \beta}\rb,
\end{eqnarray}
where $\alpha,\beta=A,B,C$ and the coefficients are given by 
 $\boldsymbol{\mathcal{M}}^0=\xi~\rm{diag}\lb 1,1,1\rb$, with $\xi=\lb h_z\cos\theta+h_x\sin\theta\rb/2+\zeta(1+\Delta)\sin^2\theta$, and the matrices $\boldsymbol{\mathcal{M}}^\pm$ differ only by a pre-factor: \begin{eqnarray}
\boldsymbol{\mathcal{M}}^\pm=-\frac{\chi \pm (1-\Delta)}{2}\boldsymbol{\Omega}
\label{com}
\end{eqnarray}
where  ${\chi}=(1+\Delta)\cos^2\theta$ and $\boldsymbol{\Omega}$  is given by
\begin{equation}
 \boldsymbol{\Omega}=\left(
\begin{array}{ccc}  
0& \cos k_1&\cos k_2 \\
\cos k_1& 0& \cos k_3\\
\cos k_2&\cos k_3 & 0 \\  
\end{array}
\right);
 \label{mat}
\end{equation} 
with  $ k_j=\bo\cdot \bold{\hat{e}}_j$; $~\bold{\hat{e}}_1 =-\lb 1/2, {\sqrt 3}/2\rb; ~ \bold{\hat{e}}_2 =\lb 1, 0\rb; ~ \bold{\hat{e}}_3 =\lb -1/2, {\sqrt 3}/2\rb$. The eigenvalues of $\boldsymbol{\Omega}$ are given by 
\begin{eqnarray}
&\omega_1=-1;\quad \omega_{2,3}=\frac{1}{2}\lb 1\pm\sqrt{1+8g_\bo}\rb;
\label{ome}
\end{eqnarray}
where  $g_\bo= \cos k_1 \cos k_2 \cos k_3$.
The Hamiltonian \ref{main} is diagonalized in two steps \cite{jon}.  Firstly,  we make a linear transformation
\begin{eqnarray}
& a_{\bo \alpha }=\sum_{\mu}\mathcal{U}_{\mu\alpha}(\bo)d_{\bo\mu},
 \label{Bogo}
\end{eqnarray}
where $\mathcal{U}_{\alpha\mu}(\bo)$  is a unitary matrix constructed from the eigenvectors of $\boldsymbol{\Omega}$ associated with the eigenvalues $\omega_\alpha$. Secondly, we apply the Bogoluibov transformation
\begin{eqnarray}
d_{\bo  \alpha }=u_{\bo \alpha }\beta_{\bo \alpha }-v_{\bo \alpha }\beta_{-\bo \alpha }^\dagger; \quad u_{\bo \alpha }^2-v_{\bo \alpha }^2=1.
 \label{Bogo1}
\end{eqnarray}
The resulting Hamiltonian is diagonalized with
\begin{eqnarray}
&u_{\bo \alpha }^2=\frac{1}{2}\lb \frac{A_{\bo \alpha }}{{\epsilon}_{\bo \alpha }}+1\rb; \quad v_{\bo \alpha }^2=\frac{1}{2}\lb \frac{A_{\bo \alpha }}{{\epsilon}_{\bo \alpha }}-1\rb,\end{eqnarray}
and
\begin{eqnarray}
&A_{\bo \alpha }=\xi+(1-\Delta)\omega_{\bo \alpha }+ B_{\bo \alpha };~ B_{\bo \alpha }=-\frac{\chi+(1-\Delta)}{2}\omega_{\bo \alpha }.\label{abex}
\end{eqnarray}
The energy is given by ${\epsilon}_{\bo\alpha}= \sqrt{A_{\bo\alpha}^2-B_{\bo\alpha}^2}.$
The diagonal Hamiltonian is given by
\begin{eqnarray}
\mathcal{H}=S\sum_{\bo,\alpha}{\epsilon}_{\bo\alpha}\lb \gamma_{\bo\alpha}^\dagger \gamma_{\bo\alpha}+\gamma_{-\bo\alpha}^\dagger \gamma_{-\bo\alpha} \rb.
\end{eqnarray}
For the triangular lattice, there is only one sublattice. We obtain
\begin{eqnarray}
&A_{\bo}=\xi+3(1-\Delta)g_{\bo } + B_{\bo };~B_{\bo }=-\frac{\chi+3(1-\Delta)}{2}g_{\bo },
\label{abex2}
\end{eqnarray}
with $\xi=\lb h_z\cos\theta+h_x\sin\theta\rb/2+3(1+\Delta)\sin^2\theta$, $\chi= 3(1+\Delta)\cos^2\theta$, and $g_{\bo}=(\cos k_x+2\cos {k_x}/{2}\cos {\sqrt{3}k_y}/{2})/3.$
The corresponding Hamiltonian  is diagonalized in a similar way, however,  without the sublattice indices.   
\subsection{Excitation spectra}
\begin{figure}[!]
\centering
\includegraphics[width=5in]{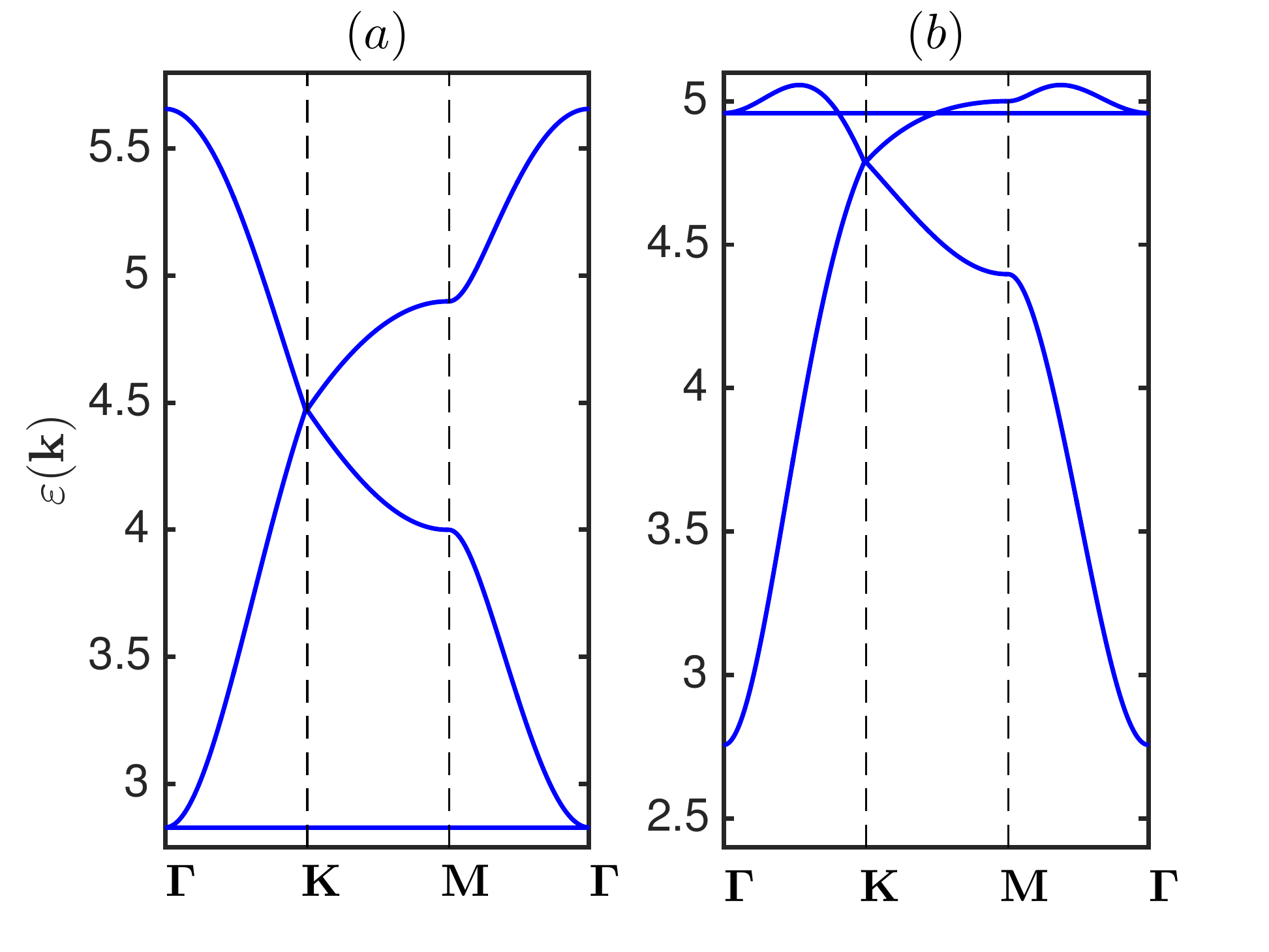}
\caption{Color online. The plots of the energy bands for the Kagom\'e lattice at $h_{x}=0$: $\Delta=0=h_z$ (a);  $\Delta=0.25;~h_z=\pm 4.5$ (b). }
\label{K_spec1}
\end{figure}

\begin{figure}[!]
\centering
\includegraphics[width=5in]{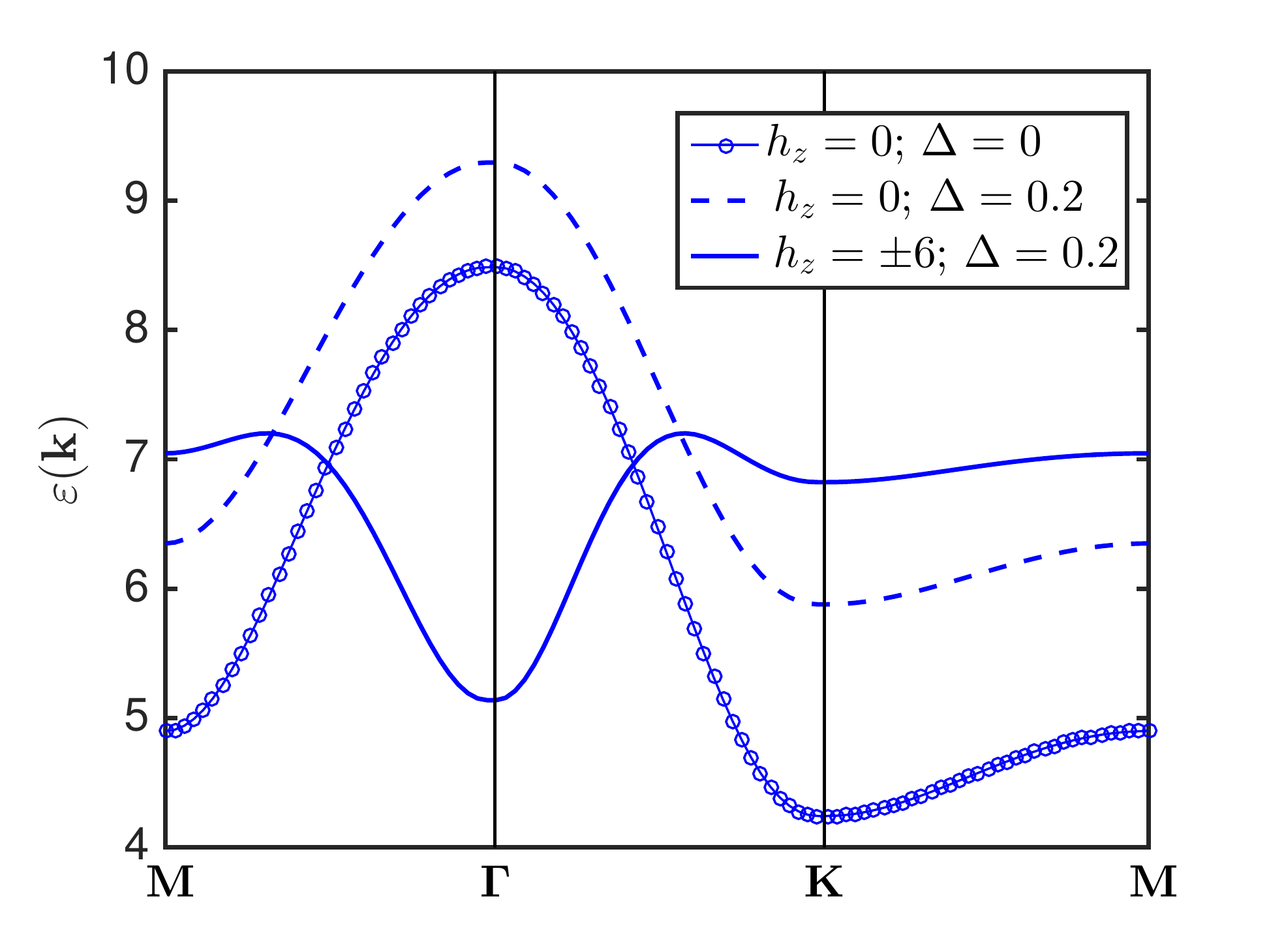}
\caption{Color online. The plot of the energy band on the  triangular lattice at $h_x=0$  and several values of $\Delta$ and $h_z$. There is only one band for each parameter.}
\label{T_spec1}
\end{figure}

The features of the quantum Kagom\'e ice hardcore bosons can be understood by  analyzing the momentum-dependent eigenfrequencies given by
$\varepsilon_{\bo\alpha}=2{\epsilon}_{\bo\alpha}$.  The energy bands for the  Kagom\'e lattice are plotted in Fig.~\ref{K_spec1}  along the Brillouin zone paths in Fig.~\ref{lattice}(b) for $\Delta=h_{x,z}=0$ (isotropic limit) and $\Delta=0.25;~h_{x}=0;~h_z=\pm 4.5$. We also   show the energy band on the triangular lattice in Fig.~\ref{T_spec1}. 
The excitation spectra of Eq.~\ref{k2}  are fully gapped on both lattices for all $\Delta$ between $0$ and $1$ for $h_{x,z}=0$. The gap persists at all values of the magnetic field $h_z\leq h_F^c$ and $h_x=0$.  The absence of a zero (soft) mode in the quantum Kagom\'e ice hardcore bosons  is a direct consequence of the discrete $Z_2$ symmetry of Eq.~\ref{k2}. For the triangular lattice, a roton minimum occurs at the corners of the Brillouin zone for $\Delta=h_{x,z}=0$, {\it i.e.,}  at $\bold{Q}_{\bf K}=(\pm 4\pi/3,0)$ and the symmetry related points, whereas the dispersion has a maximum peak at $\bold{Q}_{\bf \Gamma}=(0,0)$.   This is in stark contrast to pure U(1)-invariant XY model.
\subsection{Ground state energy}
The  spin wave correction to the mean field  energy  is given by 
\begin{eqnarray}
\Delta\mathcal{E}=\mathcal{E}_{g}-\mathcal{E}_c=S\sum_{\bo,\alpha}[{\epsilon}_{\bo\alpha}-\xi].
\label{ge}
\end{eqnarray} 
\begin{figure}[!]
\centering
\includegraphics[width=5in]{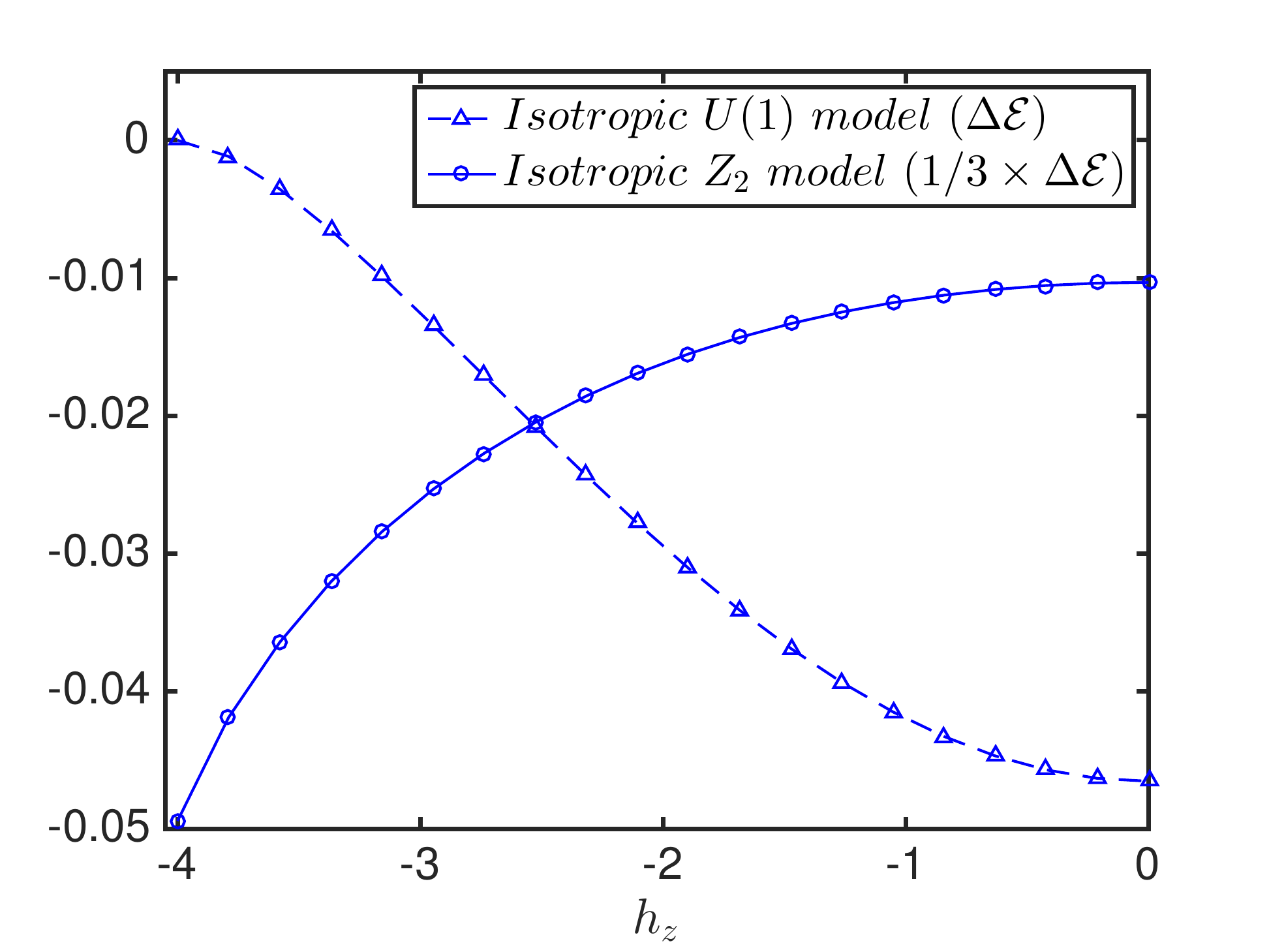}
\caption{Color online. The  spin wave zero point quantum correction as a function of $h_z$ on the Kagom\'e lattice at the isotropic $Z_2$-invariant limit $\Delta =h_{x}=0$; $S=1/2$. The U(1)-invariant model is plotted at the corresponding isotropic limit.}
\label{GS1}
\end{figure}
The correction to the ground state energy as a function of $h_z$  is plotted in  Fig.~\ref{GS1} for the Kagom\'e lattice. The triangular lattice has a similar trend. We see that the trend of the quantum Kagom\'e ice hardcore bosons is different from that of the U(1)-invariant XY model. In particular,  the energy of the isotropic $Z_2$-invariant model at $\Delta =h_{x}=0$ does not vanish at the saturated point $h_z=h_F^c$. The estimated ground state energy on the Kagom\'e lattice at the isotropic  point $\Delta =h_{z,x}=0$ is $\mathcal{E}_{g}= -0.5309$. Finite temperature QMC simulation \footref{juan1} for a relatively large cluster $V=24\times 24\times 3$ spins at inverse-temperature $\beta=100$ gives $\mathcal{E}_{g}= -0.5359(2)$, in good agreement with linear spin wave theory. For the triangular lattice, linear spin wave theory gives $\mathcal{E}_{g}= -0.7791$ \cite{sow3}; there is no QMC result in this case.

\subsection{  Particle and condensate densities}
We now study the ground state  properties of Eq.~\ref{k1} by probing the sublattice magnetizations, which will quantify the strengths of quantum fluctuations in this model. We first analyze the Kagom\'e lattice, the triangular lattice follows a similar pattern. For the Kagom\'e lattice,  the sublattice magnetizations  are equivalent, the total magnetizations per site are given by
\begin{eqnarray}
&\la S_z\ra= -\frac{1}{S\mathcal N}\frac{\partial \mathcal{E}_{SW}(h_z,h_x=0)}{\partial  h_z},\label{sz}\\&
\la S_x\ra= -\frac{1}{S\mathcal N}\left.\frac{\partial \mathcal{E}_{SW}(h_z,h_x)}{\partial  h_x}\right\vert_{h_x=0}\label{sx}.
\end{eqnarray}
Using Eq.~\ref{ge} we obtain
\begin{eqnarray}
&\la S_z\ra= S\cos\vartheta_c + \frac{\cos\vartheta_c}{2\zeta}\frac{1}{\mathcal{N}}\sum_{\bo\alpha} (\tilde{A}_ {\bo\alpha}-\tilde{B}_ {\bo\alpha}) \frac{\omega_{\bo\alpha}}{\tilde{\epsilon}_{\bo\alpha}}.
\label{sz1}
\end{eqnarray}
To linear order in $h_x$ we find
\begin{eqnarray}
&\la S_x\ra= S\sin\vartheta_c - \frac{\cos^2\vartheta_c}{2\zeta\sin\vartheta_c}\frac{1}{\mathcal{N}}\sum_{\bo\alpha}(\tilde{A}_ {\bo\alpha}-\tilde{B}_ {\bo\alpha}) \frac{\omega_{\bo\alpha}}{\tilde{\epsilon}_{\bo\alpha}}\label{sx1}\\&\nonumber
-\frac{1}{2\sin\vartheta_c}\frac{1}{\mathcal{N}}\sum_{\bo\alpha}\bigg[\frac{\tilde{A}_{\bo\alpha}}{\tilde{\epsilon}_{\bo\alpha}}-1\bigg].
\end{eqnarray}
where $\tilde{A}_{\bo\alpha}=A_{\bo\alpha}(h_x=0)$, etc. 
As seen in Eqs.~\ref{sz1} and \ref{sx1}, the denominator in these expressions  correspond to the gapped energy spectrum. We define the particle density $\rho$ and the ``condensate density'' at $\bo_\Gamma=0$ as $\rho =S+\la S_z\ra$ and $\rho_0 =\la S_x\ra^2$ respectively.  In LSWT  we find
\begin{eqnarray}
&\rho_0= (S\sin\vartheta_c)^2 - S\frac{\cos^2\vartheta_c}{\zeta}\frac{1}{\mathcal{N}}\sum_{\bo\alpha}(\tilde{A}_ {\bo\alpha}-\tilde{B}_ {\bo\alpha}) \frac{\omega_{\bo\alpha}}{\tilde{\epsilon}_{\bo\alpha}}\label{sx0}\\&\nonumber
-S\frac{1}{\mathcal{N}}\sum_{\bo\alpha}\bigg[\frac{\tilde{A}_{\bo\alpha}}{\tilde{\epsilon}_{\bo\alpha}}-1\bigg].
\end{eqnarray}
\begin{figure}[!]
\centering
\includegraphics[width=5in]{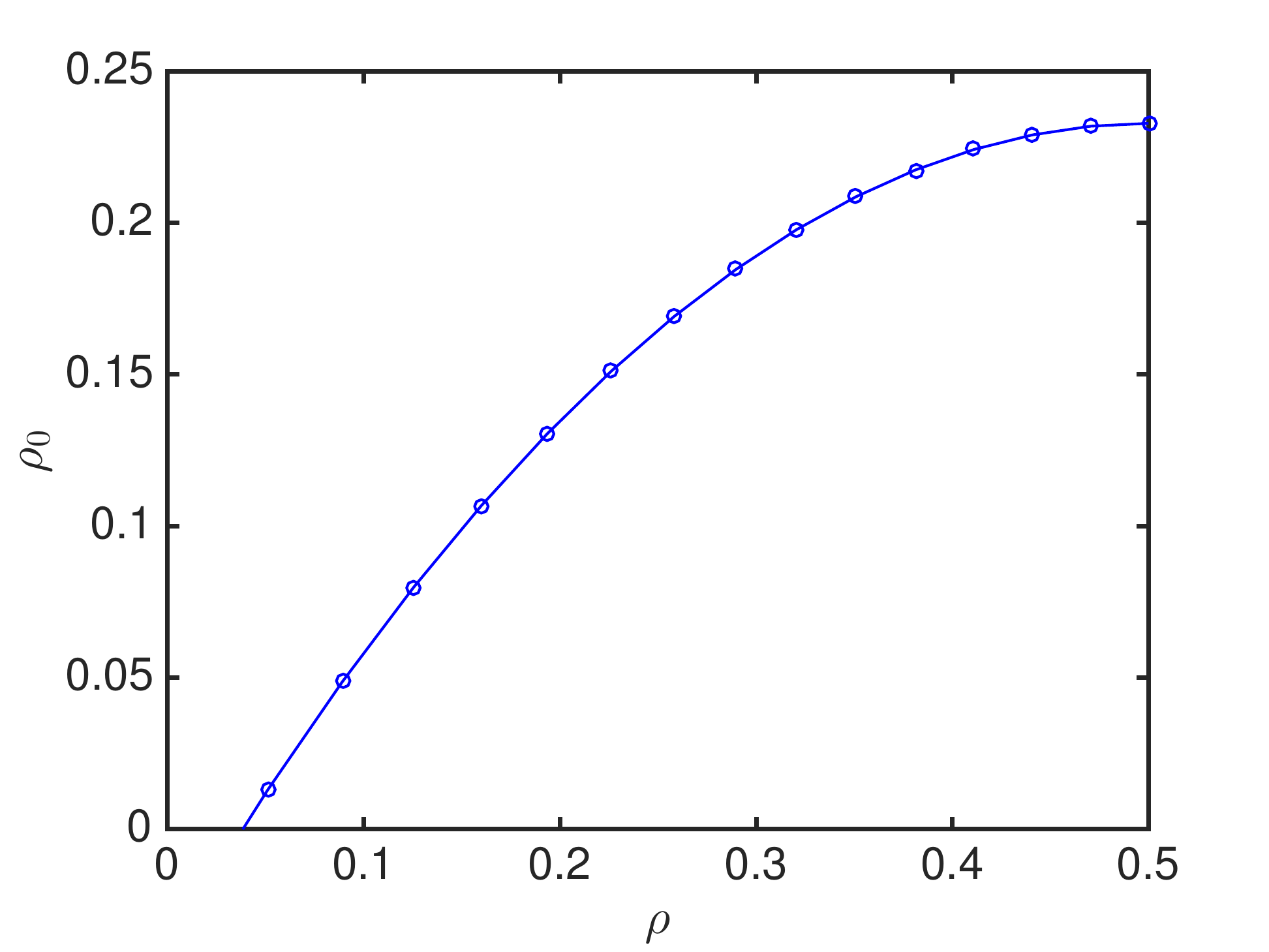}
\caption{Color online. The spin wave condensate density $\rho_0$ against the particle density $\rho$ on the  Kagom\'e lattice at $h_x=\Delta=0$ and $S=1/2$.}
\label{rho0}
\end{figure}
\begin{figure}[!]
\centering
\includegraphics[width=5in]{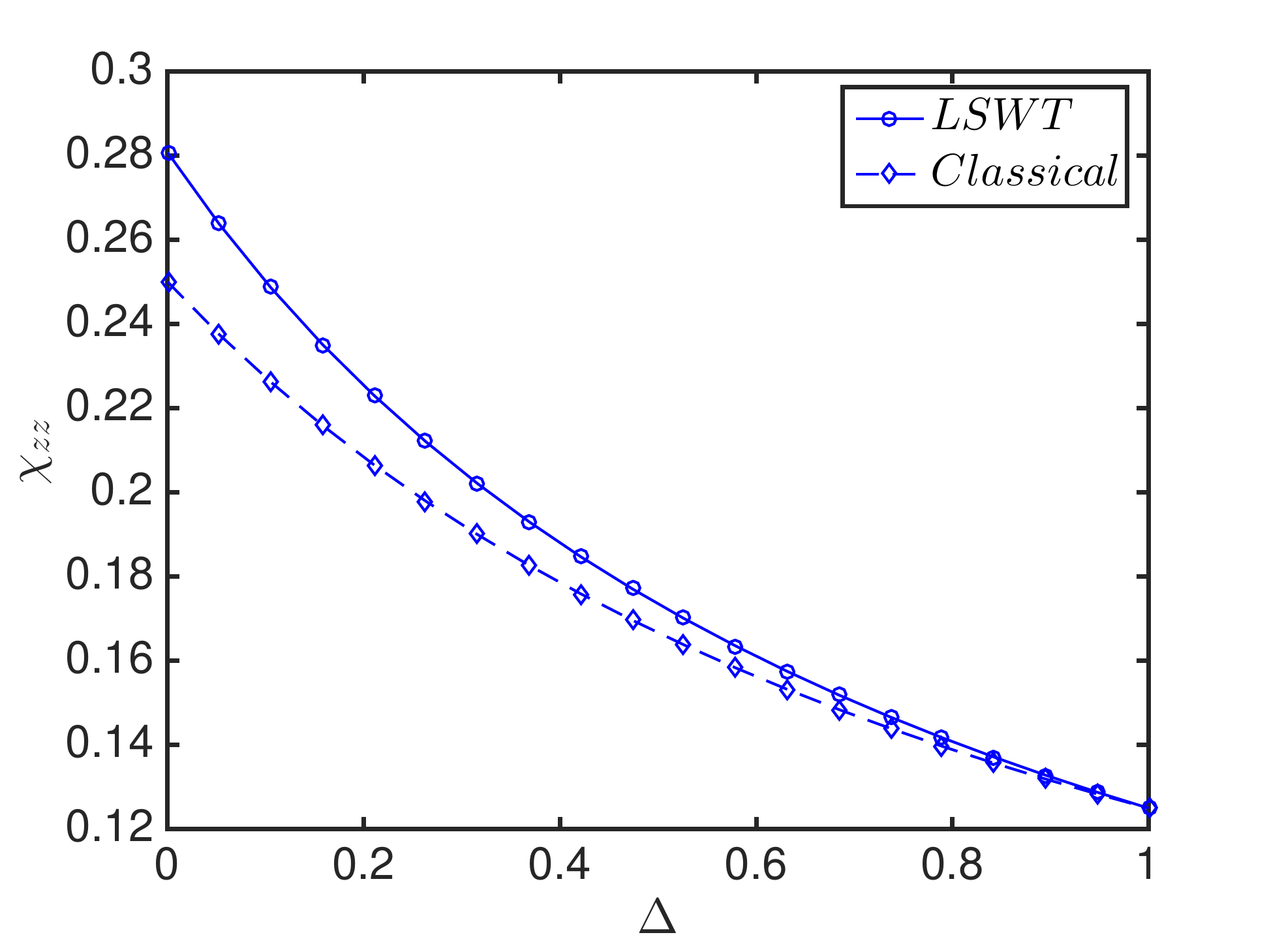}
\caption{Color online. The plot of the classical and the spin wave quantum susceptibilities as a function of $\Delta$ for the Kagom\'e lattice  at $S=1/2$.}
\label{Ksus}
\end{figure}
The corresponding expressions for the triangular lattice are similar to  the Kagom\'e lattice without the sublattice summation.  Figure~\ref{rho0} shows the plot of the condensate   against the particle density at $\Delta=0$ (isotropic $Z_2$-invariant XY model) for on the Kagom\'e lattice. At half filling ($\rho=0.5$ or $h_{x,z}=\Delta=0$), the  estimated values of the order parameter for the Kagom\'e lattice is $\la S_x\ra= 0.4829$ and $\rho_0=\la S_x\ra^2$.  Finite temperature QMC simulation \footref{juan1} for a relatively large cluster $V=24\times 24\times 3$ spins at inverse-temperature $\beta=24$ gives $\la S_x\ra= 0.4785(2)$. For the triangular lattice, linear spin wave theory gives $\la S_x\ra= 0.4902$, which is closer to the mean-field value $\la S_x\ra= 0.5000$, there is no QMC result in this case. Clearly,  the gapped   energy spectrum of Eq.~\ref{sx1}  enhances the thermodynamic quantities. 

The out-of-plane magnetic susceptibility, $\chi_{zz}$, is another important measurable quantity as it relates to the compressibility of  bosons. It is given by
\begin{eqnarray}
&\chi_{zz}= \frac{1}{S}\left.\frac{\partial\la S_z\ra}{\partial  h_z}\right\vert_{h_z=0}.
\label{zzz}\end{eqnarray}
We find
\begin{eqnarray}
&\chi_{zz}= \chi_{zz}^c + \frac{1}{4\zeta^2(1+\Delta)S}\frac{1}{\mathcal{N}}\sum_{\bo\alpha}(\bar{A}_ {\bo\alpha}-\bar{B}_ {\bo\alpha}) \frac{\omega_{\bo\alpha}}{\bar{\epsilon}_{\bo\alpha}},
\label{zz}
\end{eqnarray}
where $\chi_{zz}^c=[2\zeta(1+\Delta)]^{-1}$ is the classical susceptibility 
and $\bar{A}_{\bo\alpha}=A_{\bo\alpha}(h_{x,z}=0)$, etc. The plot of $\chi_{zz}^c$ and $\chi_{zz}$  as a function of $\Delta$  are shown in Fig.~\ref{Ksus} for the Kagom\'e lattice. The triangular lattice has a similar trend. The classical and the quantum susceptibilities coincide as $\Delta\to 1$,  corresponding to the fully polarized $S_x$-ferromagnet.  The estimated values of $\chi_{zz}$ at the isotropic $Z_2$-invariant limit, $\Delta=0$ for the Kagom\'e is $\chi_{zz}=0.2807$.  Finite temperature QMC simulation \footref{juan1} for a large cluster $V=24\times 24\times 3$ spins at inverse-temperature $\beta=24$ gives $\chi_{zz}= 0.2785(2)$. For the triangular lattice, linear spin wave theory gives  $\chi_{zz}=0.1799$, and the mean-field value $\chi_{zz}= 1/6$; again there is no QMC result in this case.  In contrast to the U(1)-invariant model \cite{ber, tom, B,san, fu,fu1, gom, zhe,zhe1},  we see that quantum fluctuation increases the classical susceptibility  away from $\Delta=1$.
\subsection{  Dynamical structure factors}
Furthermore, we explore the  nature of the  quantum Kagom\'e ice hardcore bosons  by studying  the dynamical structure factor, which is an important quantity in experiments for characterizing the ground state properties of a quantum system. The spin structure factor is given by the Fourier transform of the equal-time spin-spin correlation function
\begin{eqnarray}
\mathcal{S}^{\mu\nu}(\bo)= \frac{1}{\mathcal{N}}\sum_{l,m} e^{i\bo\cdot (\bold{r}_{l}-\bold{r}_{m})}\la S_{l}^\mu S_{m}^\nu\ra,
\end{eqnarray}
where  $\mu,\nu=(x,y,z)$ label the components of the spins and $S_{l}^\mu=\sum_\alpha S_{l\alpha}^\mu$ on the Kagom\'e lattice,  $\alpha$ labels the sublattices. We restrict the calculation of the structure factors  to the case of  half-filling ($\rho=0.5$)  or zero magnetic fields ( $h_{x,z}=0$). 
 
  The momentum distribution is related to the  off-diagonal static structure factor $\mathcal{S}^{\pm}(\bo)$. We have $\la S_l^+S_m^-\ra= \la S_l^{x}S_m^{ x}\ra + \la S_l ^{ y} S_m ^{y}\ra+\la S_l ^{z}\ra \delta_{lm}$. At $h_{x,z}=0$,  $\theta=\pi/2$, hence rotation about the $y$-axis gives  $S_l^x\to S_l^{\prime z},~S_l^z\to -S_l^{\prime x},$ and $S_l^y\to S_l^{\prime y}$. The off-diagonal structure factor in the rotated coordinate up to order $S$ is given by $\mathcal{S}^{\pm}(\bo)= \la S_\bo^{\prime z}S_{-\bo}^{\prime z}\ra +  \la S_\bo^{\prime y}S_{-\bo}^{\prime y}\ra.$   At  $\bo_{\bf \Gamma}=0$,  $\mathcal{S}^{\pm}(\bo)$  is related to the order parameter  by $\rho_0=\mathcal{S}^{\pm}(\bo_\Gamma)/\mathcal{N}$ as $\mathcal{N}\to \infty$.
 \begin{figure}[!]
\centering 
\includegraphics[width=6in]{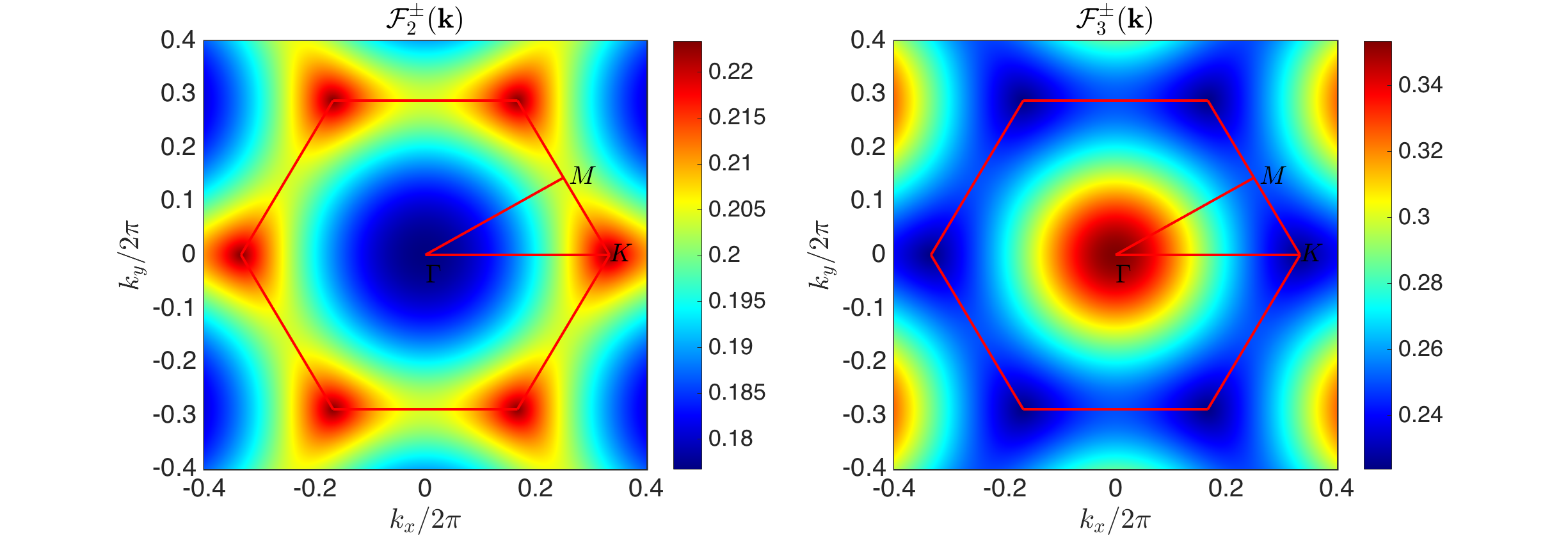}
\caption{Color online.  The  plot of the off-diagonal  form factor $\mathcal{F}^{\pm}(\bo)$ for the two dispersive bands on the Kagom\'e lattice at $S=1/2$;  $\Delta=h_{x,z}=0$.}
\label{Kpm}
\end{figure}
\begin{figure}[!]
\centering
\includegraphics[width=6in]{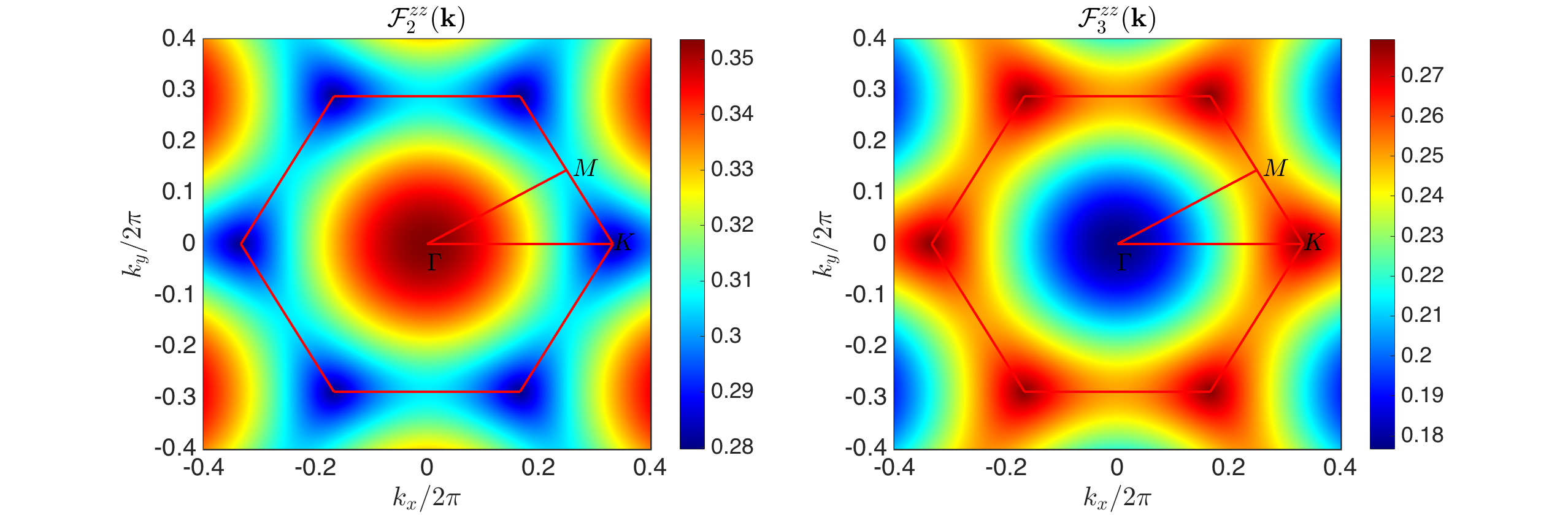}
\caption{Color online.  The plot of the out-of-plane form factor $\mathcal{F}^{zz}(\bo)$ for the two dispersive bands on the Kagom\'e lattice at $S=1/2$;  $\Delta=h_{x,z}=0$. }
\label{Ksy}
\end{figure}
 \begin{figure}[!]
\centering 
\includegraphics[width=6in]{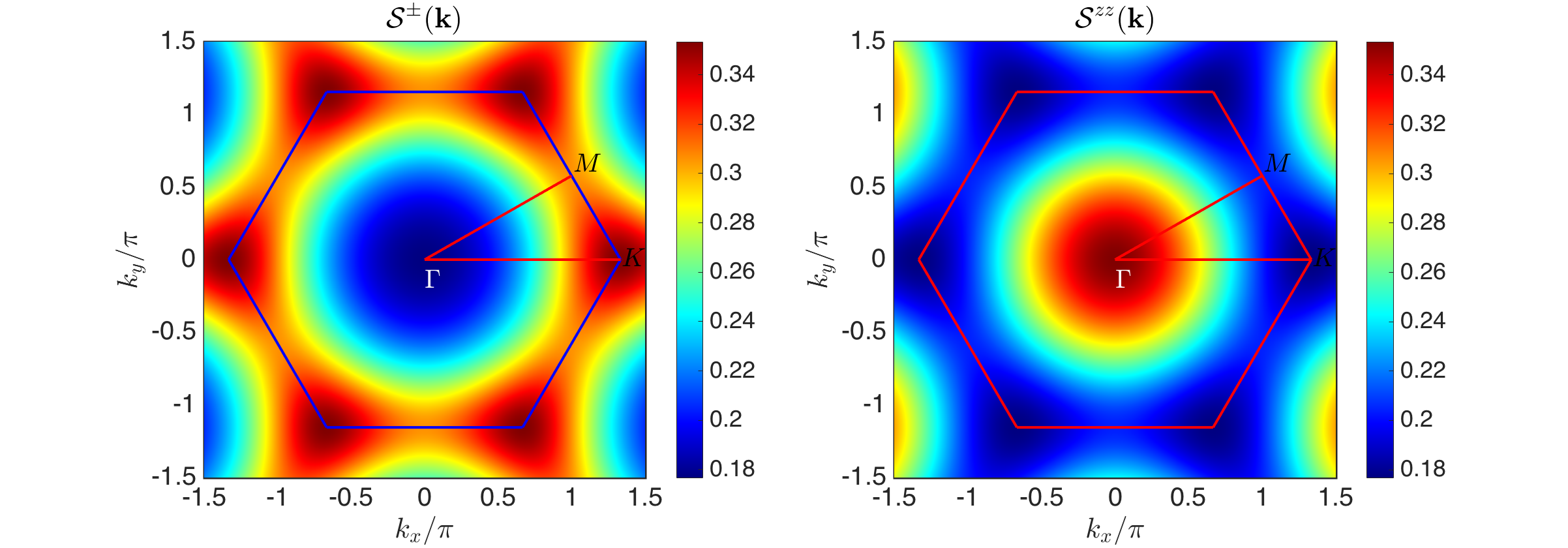}
\caption{Color online.  The  plot of the off-diagonal  structure factor $\mathcal{S}^{\pm}(\bo)$ and out-of-plane structure factor $\mathcal{S}^{zz}(\bo)$ for the triangular lattice at $S=1/2$;  $\Delta=h_{x,z}=0$.}
\label{Tpm}
\end{figure}
\begin{figure}[!]
\centering
\includegraphics[width=5in]{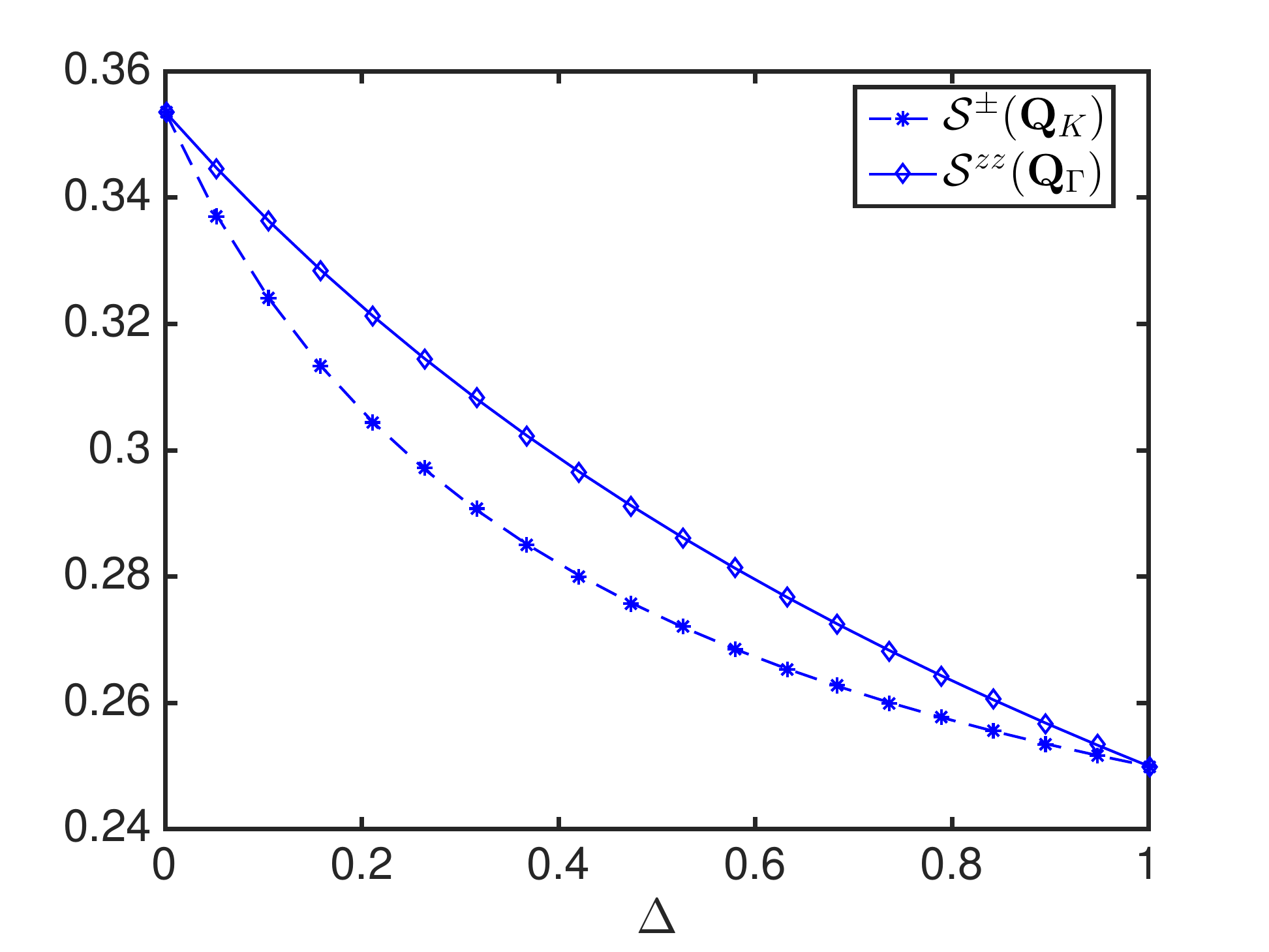}
\caption{Color online.  The plot of the structure factors in Eqs.~\ref{qk} and \ref{q0} as a function of $\Delta$ at $h_{x,z}=0$, $S=1/2$. }
\label{TQ}
\end{figure}
For the Kagom\'e lattice we find that the off-diagonal structure factor for $\bo\neq \bo_\Gamma$ and the out-of-plane structure factor are given by \bea \mathcal{S}^{\pm}(\bo)= \sum_\mu \mathcal{F}_{\mu}^{\pm}(\bo),~~\mathcal{S}^{zz}(\bo)= \sum_\mu \mathcal{F}_{\mu}^{zz}(\bo),\eea where $\mathcal{F}_{\mu}^{\pm}(\bo)= S(u_{\mu\bo} +v_{\mu\bo})^2/{2}\sum_{\alpha\alpha^\prime}\mathcal{U}_{\mu\alpha}(\bo)\mathcal{U}_{\mu\alpha^\prime}(\bo)$ and $\mathcal{F}_{\mu}^{zz}(\bo)= S(u_{\mu\bo} -v_{\mu\bo})^2/{2}\sum_{\alpha\alpha^\prime}\mathcal{U}_{\mu\alpha}(\bo)\mathcal{U}_{\mu\alpha^\prime}(\bo)$.  For the triangular lattice, the explicit expressions are
 
\begin{eqnarray}
\mathcal{S}^{\pm}(\bo)=\frac{S}{2}(u_{\bo}+v_{\bo})^2= \frac{S}{2}\lb \frac{A_{\bo}+B_\bo}{\sqrt{A_{\bo}^2-B_\bo^2}}\rb,\label{spm}\\\mathcal{S}^{zz}(\bo)= \frac{S}{2}(u_{\bo} -v_{\bo})^2=\frac{S}{2}\lb \frac{A_{\bo}-B_\bo}{\sqrt{A_{\bo}^2-B_\bo^2}}\rb 
\label{sz}.
\end{eqnarray}


The denominator of Eqs.~\ref{spm} and \ref{sz} correspond to the energy spectrum, which is gapped in the entire Brillouin zone.  In Figs.~\ref{Kpm} and \ref{Ksy} we have shown the off-diagonal  form factor $\mathcal{F}^{\pm}(\bo)$  and the out-of-plane form factor $\mathcal{F}^{zz}(\bo)$ for the two dispersive bands on the Kagom\'e lattice. Figure~\ref{Tpm} shows the off-diagonal and the out-of-plane structure factors on the  triangular lattice. The solid hexagons denote the respective Brillouin zones for the two lattices. For the Kagom\'e lattice, the momentum distribution shows noticeable non-divergent sharp peaks at the center and corners of the Brillouin zones, whereas for the triangular lattice we observe non-divergent peaks at $\bold{Q}_{\bf K}=(\pm 4\pi/3,0)$ and symmetry related points. This is due to the roton minima of the excitation spectrum in at $\bold{Q}_{\bf K}=(\pm 4\pi/3,0)$  as opposed  to a divergent peak at $\bo_{\bf \Gamma}=0$ in the U(1)-invariant model \cite{ber, san, tom,gom}.   On the other hand, the out-of-plane structure factor, $\mathcal{S}^{zz}(\bo)$, shows a non-divergent sharp peak  at $\bo_{\bf \Gamma}=0$. At  $\bold{Q}_{\bf K}=(\pm 4\pi/3,0)$ and $\bold{Q}_{\bf \Gamma}=(0,0)$ we have
 \begin{eqnarray}
 &\mathcal{S}^{\pm}(\bold{Q}_{\bf K})=S\sqrt{\frac{1+\Delta}{2(1+3\Delta)}},\label{qk}\\&  \mathcal{S}^{zz}(\bold{Q}_{\bf \Gamma})=\frac{S}{\sqrt{2(1+\Delta)}}.
 \label{q0}
 \end{eqnarray}
Figure~\ref{TQ} shows the trend of $\mathcal{S}^{\pm}(\bold{Q}_{\bf K})$ and $\mathcal{S}^{zz}(\bold{Q}_{\bf \Gamma})$ as a function of $\Delta$.

\section{ Conclusion}
 We have shown that linear spin wave theory works remarkably well in the description of the quantum Kagom\'e ice hardcore bosons. Due to the $Z_2$-invariance of the  quantum Kagom\'e ice hardcore boson model, the excitation spectra is fully gapped at all momenta and  the resulting occupation number is very small. The gapped nature of the excitation spectrum enhances the estimated values of the thermodynamic quantities. We showed that these values are in very good agreement with  finite temperature quantum Monte Carlo (QMC) simulations. We observed Bragg peaks in the structure factors at the center and corners of the Brillouin zone for the Kagom\'e lattice. This is a special feature of the $Z_2$-invariance of the Hamiltonian. For the triangular lattice, we also observed  Bragg peaks  in the structure factors at the corners of the Brillouin zone.   These are consequences of the roton minimum in the energy spectrum.  Although we have partially studied this model  numerically, the results presented in this paper are sufficient to understand the nature of the quantum Kagom\'e ice hardcore bosons.  However,  it would be interesting to explore other properties of this model by an explicit zero temperature QMC analysis.  
\section*{ Acknowledgments}
The author  would like to thank Juan Carrasquilla for providing him with the QMC results. This work was conducted at African Institute for Mathematical Sciences (AIMS).  Research at Perimeter Institute is supported by the Government of Canada through Industry Canada and by the Province of Ontario through the Ministry of Research
and Innovation.

\end{document}